\documentclass[11pt,amstex]{article}

\usepackage{amsmath}
\usepackage{amssymb}

\numberwithin{equation}{section}

\begin{document}
\title{Quantum Mechanics in a Rotating Frame}

\author{Jeeva Anandan\thanks{jeeva@sc.edu} \ and Jun Suzuki\thanks{suzuki@physics.sc.edu} \\
\it Department of Physics and Astronomy, \\ \it University of South Carolina, Columbia SC 29208 } 
\date{}


\maketitle

\begin{abstract}
The rotating frame is considered in quantum mechanics on the basis of the position dependent boost relating this frame to
the non rotating inertial frame. 
We derive the Sagnac phase shift and the spin coupling with the rotation in the non relativistic limit by a simple treatment. 
By taking the low energy limit of the Dirac equation with a spin connection, we obtain the Hamiltonian for the rotating frame, which 
gives rise to all the phase shifts discussed before. Furthermore, we obtain a new phase shift due to the spin-orbit coupling.
\end{abstract}

\section{Introduction} 

The rotating frame has played an important role both in classical and quantum physics. One reason for this is that
thermal equilibrium in a closed system can be realized when  the system has uniform translation and 
rotation relative to an inertial frame \cite{landau}.  
So, the macroscopic properties of the system are not affected by the uniform rotation, apart from the influence of centrifugal and  Coriolis fields.
This result is significant because most experiments are  done under the influence of the earth's rotation. 
In a quantum system, there are also global consequences of rotation, such as the phase shift in interferometry (Sagnac effect).
An experiment to detect  the Sagnac effect due to the earth's rotation in neutron interferometry by using  a vertical  incoming beam was proposed  by one of  us \cite{an1977}, which led to this experiment being performed subsequently by Werner, Staudenmann, and Colella \cite{we1979}.
There were many discussions of this effect in the past three decades \cite{a}, which we are in agreement with.
However there still remain misconceptions, which may be a source of confusion for some people.
Therefore we consider it as a good opportunity to clarify 
the consequences of quantum mechanics when it is applied to a rotating frame.

This paper continues as follows:
in section 2, we  begin with the Lorentz boost in special relativity,  and  by taking its non relativistic limit  obtain all possible  ways of 
implementing the Galilei boost in  quantum mechanics. 
Then in section 3, we  discuss a rotating frame in non relativistic quantum mechanics and 
obtain the Hamiltonian and derive phase shifts. We, furthermore, discuss relativistic aspects of the rotating frame to understand 
the limitation of non relativistic approach, in section 4 and obtain a Hamiltonian in the low energy limit of the Dirac equation.
 All the phase shifts in a rotating frame, including a new phase shift due to the spin-orbit coupling, are then obtained from this Hamiltonian. 
We use $\hbar = c =1$ units throughout the paper unless we write them explicitly.


\section{Lorentz and Galilei transformations in Quantum Mechanics} 

We shall consider a spinless particle to make our discussion clear and construct all possible Galilei transformations in the non relativistic limit.
The scalar field $\phi (x^{\mu})$ is transformed under the infinitesimal Lorentz boost 
$\Lambda ^{\mu} _{\ \nu} \simeq I ^{\mu} _{\ \nu}  + \omega ^{\mu} _{\ \nu} $,
\begin{equation}
\delta \phi (x^{\mu}) = \phi (x^{'\mu} =\Lambda ^{\mu} _{\ \nu} x^{\nu}) - \phi (x^{\mu}) \simeq \frac i 2 \omega _{\mu \nu } L^{\mu \nu} \phi (x^{\mu})
\end{equation}
where $L^{\mu \nu}$ are the generators of the Lorentz transformations defined by
\begin{equation}
L^{\mu \nu} =  x^{\mu} p^{\nu} - x^{\nu} p^{\mu} . \label{L}
\end{equation}
Threfore the infinitesimal Lorentz boost for the direction $x^i$ is given by
\begin{equation}
U  \simeq  I +  i \omega _{0 i} L^{0i} ,
\end{equation}
and hence, we obtain the Lorentz boost $U$ as
\begin{equation}
 U = \exp ( i \omega _{0 i}(x^0 p^i - x^i p^0)) .
\end{equation}
Now we take the non relativistic limit for the above boost and it becomes the Galilei boost with a velocity $\bf V$
\begin{equation}
U = \exp (i t {\bf V} \cdot \hat{ \boldsymbol p} - i m {\bf V} \cdot \hat{ \boldsymbol x} ) . \label{u}
\end{equation}

Using the Galilei boost (\ref{u}), we can implement a boost from one inertial frame $F_0$ 
and another inertial frame $F'_0$ which is related to $F_0$ by the velocity $\bf V$, i.e. 
\begin{equation}
\boldsymbol x' = \boldsymbol x - {\bf V} t , \quad t' = t 
\end{equation}
where  unprime quantities refer to the frame $F_0$ and prime quantities refer to the frame $F'_0$. 
There are two natural and equivalent ways to do it as shown below. \\[5pt]
i) $U$ acting on the wave function; 
\begin{eqnarray}
\psi ' ( \boldsymbol x' ,t') &=& \exp (- i m {\bf V} \cdot \boldsymbol x + i \frac 12 m {\rm V}^2 t) \psi (\boldsymbol x ,t)  \label{psi1} \\
\hat{\boldsymbol p'} \quad &=&\hat{ \boldsymbol p } \label{p1}
\end{eqnarray}
\noindent ii) $U$ acting on the momentum operator;
\begin{eqnarray}
\psi ' ( \boldsymbol x' ,t') &=&  \psi (\boldsymbol x ,t) \label{psi2} \\
\hat{\boldsymbol p' } \quad &=& U^{\dagger} \hat{\boldsymbol p} U = \hat{ \boldsymbol p} - m {\bf V} \label{p2}
\end{eqnarray}

The equivalence between i) and ii) is checked easily, if we notice that there exist a local gauge transformation 
between two pictures such as,
\begin{equation}
e^{i f(\boldsymbol x ,t)} = \exp (- i m {\bf V} \cdot \boldsymbol x + i \frac 12 m {\rm V}^2 t) .
\end{equation}
And we can immediately see the equivalence by rewriting i) as $\psi ' = e^{if} \psi$, $\hat{ \boldsymbol p'} =  \hat{ \boldsymbol p} $, 
and ii) as $\psi ' =  \psi$, $\hat{ \boldsymbol p'} =  e^{-if}\hat{ \boldsymbol p} e^{if}$. 
This is like the difference between the Schr\"odinger picture; i) and the Heisenberg picture; ii).  And 
it should be emphasized that observed quantities are neither operators nor the wave functions themselves but expectation values which are 
calculated from them, and (of course) two pictures yield same expectation values. 
Although above two methods seem natural to transform one frame to another, there are also an infinite number of ways which 
are related to above methods by some other gauge transformations.  And this exhausts all possible ways of implementing 
the Galilei boost in quantum mechanics.

Before we discuss the Hamiltonian of the system, let us consider a non trivial example which helps us understand 
the physics behind those two pictures. Suppose the wave function in the frame $F_0$ is given by a plane wave $e^{i k x} $
($k=2 \pi / \lambda$), and we examine the wave function seen from the frame $F'_0$.  In the picture i) the wave length is changed 
due to the phase factor in front of (\ref{psi1}), on the other hand in the picture ii) the wave length does not change since the wave 
function transforms as a scalar.  However the momentum operator does change in ii) as (\ref{p2}), and this is consistent 
with the fact that the de Broglie relation $p=h / \lambda$ holds in all frames.

Next we shall examine the Schr\"odinger equation of the system and discuss energies measured in both frames.
As we already saw, two methods i) and ii) are equivalent. Therefore it is enough to examine ii) only.
Starting with the Schr\"odinger equation in the frame $F_0$
\begin{equation}
i \frac {\partial }{\partial t} \psi (\boldsymbol x ,t) = H \psi (\boldsymbol x ,t), \quad H = \frac{\hat{ \boldsymbol p}^2}{2m} , \label{sch1}
\end{equation}
we can transform it to the frame $F'_0$ as
\begin{equation}
i (\frac {\partial }{\partial t'} + {\bf V} \cdot \boldsymbol \nabla ' ) \psi (\boldsymbol x' ,t') =
 H' \psi (\boldsymbol x' ,t'),  \label{sch2}
\end{equation}
where the left hand side is obtained by the chain rule and $H'$ is 
\begin{equation}
H' = U^{\dagger} H U = \frac{(\hat{ \boldsymbol p} - m {\bf V})^2}{2m} = \frac{\hat{ \boldsymbol p'}^2}{2m} . \label{H2}
\end{equation}
So we identify the energy operator $\hat{E'}$ in $F'$ as $\hat{E'}=i (\frac {\partial }{\partial t'} + {\bf V} \cdot \boldsymbol \nabla ' )$ 
whose eigenvalue is positive. Notice that the energy measured in the frame $F'_0$ is also obtained from the non relativistic limit of 
the Lorentz transformation for the energy, namely,
\begin{equation}
E' = (1- {\rm V}^2)^{- \frac 12} (E -  {\bf V} \cdot \boldsymbol p) = E -  {\bf V} \cdot \boldsymbol p + \frac 12 m {\rm V}^2 
\end{equation}
which agrees with the result (\ref{H2}).
Now we rewrite the equation (\ref{sch2}) in the following form using $\hat{ \boldsymbol p'} =\hat{ \boldsymbol \nabla '} - m {\bf V}$,
\begin{equation}
(i \frac {\partial }{\partial t'} +\frac 12 m {\rm V}^2) \psi (\boldsymbol x' ,t') = \frac{1}{2m}(\hat{ \boldsymbol p'} - m {\bf V})^2  \psi (\boldsymbol x' ,t') .  \label{sch3}
\end{equation}
Therefore we find that this result is equivalent as the one obtained from the minimal coupling with a gauge field ${\cal A}^{\mu} =( - \frac 12  {\rm V} ^2 , {\bf V})$ 
by a coupling constant $m$, namely,
\begin{equation}
\hat{  p^{\mu}} \rightarrow \hat{  p^{\mu}} - m {\cal A}^{\mu} .
\end{equation}

\section{Non Relativistic Aspects of the Rotating Frame} 

As is shown in the previous section there is no difficulty to implement the Galilei boost in quantum mechanics, 
next we shall extend the above method to a rotating frame $F'$ whose angular velocity with respect to $F_0$ is 
$\boldsymbol \Omega$ (a constant of the time; we are dealing with a uniform rotating system throughout the paper.). 
One may try to construct the boost for the rotating frame in the following way which leads a shortcoming.  
Since the velocity $\bf V$ is, now, given by $\boldsymbol \Omega \times \boldsymbol x$, the substitution it into (\ref{u}) gives 
the boost $U =\exp (it (\boldsymbol \Omega \times \hat{\boldsymbol x}) \cdot \hat{\boldsymbol p}) =  
\exp (i t \boldsymbol \Omega \cdot \hat{\boldsymbol L})$ where $\hat{\boldsymbol L} = \hat{\boldsymbol x} \times \hat{\boldsymbol p}$ 
is the orbital angular momentum operator of the particle. So one might conclude that 
the wave function transforms as a pure rotation from $F_0$ to $F'$ in the picture i), or since $U$ commutes with $ \hat{\boldsymbol L}$
one might predict that the orbital angular momentum is same in both frames.  
However those consequences are obviously wrong even classically, the reason is due to the fact that the boost 
$U= e^{i t \boldsymbol \Omega \cdot \hat{\boldsymbol L} }$ transforms from $F_0$ to $F'_0$, but not to $F'$.

To resolve this shortcoming we need to realize that the boost from $F_0$ to $F'$ depends on the position and therefore it cannot be 
expressed as a single transformation. In general, two successive Lorentz transformations are written as the product of the Lorentz transformation and the rotation 
and hence, the boost, in this case, cannot be like a simple form as the one (\ref{u}) obtained before. 
The easiest way to get the correct result in the non relativistic limit is the minimal coupling with the gauge field as is 
mentioned before. The gauge field ${\cal A}^{\mu}$ for the rotating frame is, now, defined by
\begin{equation}
{\cal A}^{\mu} (x^{\mu}) =({\cal A}_0(\boldsymbol x) ,\boldsymbol {\cal A} (\boldsymbol x)) 
= ( - \frac 12  ( \boldsymbol \Omega \times \boldsymbol x) ^2 , \boldsymbol \Omega \times \boldsymbol x ) . \label{bmu}
\end{equation}
And we obtain the Hamiltonian for a particle at rest with respect to the rotating frame $F'$ as 
\begin{equation}
H = \frac{1}{2m}(\hat{ \boldsymbol p} - m \boldsymbol \Omega \times  \boldsymbol x )^2 
- \frac{1}{2} m ( \boldsymbol \Omega \times \boldsymbol x )^2 , \label{H3} 
\end{equation}
where we drop primes under the understanding.
Notice that one can obtain the semiclassical equation of motion for the expectation value using the Heisenberg equation of motion, i.e. 
\begin{equation}
m  \frac{ d^2  \langle \boldsymbol x \rangle }{d t^2} 
 =  2 m \frac{d \langle \boldsymbol x \rangle}{dt} \times \boldsymbol \Omega 
+m \boldsymbol \Omega \times ( \langle \boldsymbol x \rangle \times \boldsymbol \Omega) 
\end{equation}
which recover the Coriolis force and the centrifugal force correctly. 

In order to take into account the spin of the particle (we consider the neutron, namely spin $\frac 12$ particle here.), 
we need to realize the fact that the spin in $F'$ rotates with the angular velocity $- \boldsymbol \Omega$ 
relative to the inertial frame $F_0$. So the interaction between the spin and the rotation is simply the same as 
the Thomas precession \cite{a3,a2}, therefore the Hamiltoninan is obtained by adding the spin interaction term to (\ref{H3}), 
\begin{equation}
H = \frac{1}{2m}(\hat{\boldsymbol p} - m \boldsymbol \Omega \times  \boldsymbol x )^2
- \frac{1}{2} m ( \boldsymbol \Omega \times \boldsymbol x )^2  -\boldsymbol \Omega  \cdot \hat{\boldsymbol S} . \label{H4} 
\end{equation}

Thus we derive the Sagnac phase shift from (\ref{H4}) in the same manner as the Aharonov-Bohm effect \cite{ac,semon,sakurai};
\begin{equation}
\delta \phi _{\rm Sagnac} = \frac {m}{\hbar} \oint  d \boldsymbol l \cdot ( \boldsymbol \Omega \times \boldsymbol x) 
= \frac {2 m}{\hbar} \int d \boldsymbol s \cdot  \boldsymbol \Omega = \frac {2 m \boldsymbol A \cdot \boldsymbol \Omega }{\hbar} ,
\end{equation}
where $\boldsymbol A$ is the orientated area enclosed by the path of the neutron beam.
The effect of coupling of the spin to the rotation (the last term in (\ref{H4})) is to act on the initial wave function by the operator \cite{mashhoon,a3}
\begin{equation}
\hat{\Phi}_{Spin} = \hat{T} [ \exp (\frac{i}{\hbar} \int dt \hat{ \boldsymbol S} \cdot  \boldsymbol \Omega)],
\end{equation}
where $\hat{T}$ represents the time ordering operator. For the uniform rotational frame it is reduced to 
\begin{equation}
\hat{\Phi}_{Spin} = \exp (\frac{i}{\hbar} \hat{ \boldsymbol S} \cdot  \boldsymbol \Omega t)
=\hat{I} \cos (\frac{\Omega t}{2}) + i \frac{\hat{\boldsymbol \sigma} \cdot  \boldsymbol \Omega}{|\boldsymbol \Omega|} \sin(\frac{\Omega t}{2}),
\end{equation}
where $\hat{\boldsymbol \sigma} = (\sigma _x \; \sigma _y \; \sigma _z)$ are the Pauli spin matrices.
This phase shift can be observed in nuclear or molecular beam resonance methods.


\section{Relativistic Aspects of the Rotating Frame} 

In this section we shall discuss the relativistic aspects of the rotating frame in quantum mechanics, using the Dirac equation with a spin connection;
\begin{equation}
(i \gamma ^{\mu} \nabla _{\mu} -m) \psi =0 \label{dirac1}
\end{equation}
\begin{equation}
\nabla _{\mu} = \partial _{\mu} - \frac i4 \Gamma ^{ab} _{\  \mu} M_{ab}, \ M^{ab}=\frac i2 [\gamma ^a , \gamma ^b] ,\label{cov}
\end{equation}
where we summarize conventions and notations in the appendix.
The metric in a uniform rotating frame is 
\begin{equation}
g_{00} =1 -( \boldsymbol \Omega \times \boldsymbol x) ^2,  \; g_{ii}= -1, \; g_{0i}=- (\boldsymbol \Omega \times \boldsymbol x)^i ,(i=1,2,3),
\end{equation}
and $g_{\mu \nu}=0$ otherwise \cite{landau2}, then the problem becomes the simpler and solvable in the low energy limit. 
Using the properties of gamma matrices and the vierbein in the appendix, rewriting the spinors as
$\psi \rightarrow e^{-imt} \psi$, and neglecting terms of order $v^2/c^2$, where $v$ is the velocity {\it relative to the rotating frame}
, we obtained the low energy limit of the Dirac equation (\ref{dirac1}) as
\begin{equation}
[\gamma ^0 (m + \hat{p} _0 -m {\cal A}_0 - \frac 12 {\cal A}^i \hat{p}_i)
+ \gamma ^i (\hat{p}_i - \frac 12 m {\cal A}_i -\frac i2 {\cal E}_i) -m] \psi =0, \label{dirac2}
\end{equation}
where $\boldsymbol {\cal E}$ is an analog of the electric field and is defined by 
\begin{equation}
\boldsymbol {\cal E} = - \frac 12 \boldsymbol \nabla h_{00} = - \boldsymbol \nabla {\cal A}_0.
\end{equation}
Then, using the usual splitting of the four spinors into upper and lower components
 the Hamiltonian for the upper component two spinors in the low energy limit is 
\begin{equation}
H = \frac{1}{2m}(\hat{\boldsymbol p} - m \boldsymbol {\cal A} - \hat{\boldsymbol S} \times \boldsymbol {\cal E} )^2
+m {\cal A} _0 - \boldsymbol \Omega  \cdot \hat{\boldsymbol S} . \label{H5}
\end{equation}
There is an additional term ($-\frac{1}{8m}  \boldsymbol \nabla \cdot  \boldsymbol {\cal E}$) which is an analogous to 
the darwin term in the electromagnetic field case. In the present case this term is $- \frac{3 \Omega ^2}{8m c^2}$  
which is a constant, therefore it can be subtracted away from the Hamiltonian (\ref{H5}). 
In the rotating frame $\boldsymbol {\cal E} = \Omega ^2 \boldsymbol x/c^2$ so if we now neglect 
$\hat{\boldsymbol S} \times \boldsymbol {\cal E}$ term which is of order $c^{-2}$ then we obtain the Hamiltonian (\ref{H4})  \cite{com2}. 
We obtain not only the phase shifts discussed in section 3, but also we get a new phase shift 
that is calculated by acting on the initial wave function by the following operator, 
\begin{equation}
\hat{\Phi}  =  P [ \exp (\frac {i}{ \hbar} \oint  d \boldsymbol l \cdot ( \hat{ \boldsymbol S } \times \boldsymbol {\cal E})) ]
\end{equation}
where $P$ denotes the path ordering. As a special ideal case we consider, for a simplicity, a circular path. Then 
\begin{equation}
\hat{\Phi}  =\exp (\frac {i2 \Omega ^2}{ \hbar c^2}  \boldsymbol A \cdot  \hat{\boldsymbol S}).
\end{equation}
Moreover, if the spin is polarized perpendicular to the plane of the interferometry, the phase shift $\delta \phi$ due to this operator is,
\begin{equation}
\delta \phi = \frac { \Omega ^2 A }{ c^2}  .
\end{equation}
This phase shift is analogous to the phase shift due to the electric field in neutron interferometry found by Anandan \cite{a6}, Aharonov and Casher \cite{a-c}.
 Although this phase shift is very small compared to the dominant Sagnac term,
 it is interesting because it is due to a new spin-orbit coupling, and we hope that it would be experimentally tested in the future.


\section{Conclusion} 

We have shown that the significance of boosts in the treatment of the rotating frame in quantum mechanics.
And also it is suggested \cite{a2,ac,semon,sakurai,a1} that the rotating frame is considered like a gauge field in the non relativistic limit, 
 However, as we discussed in section 4, it should be remarked that there do exist several differences 
between the gauge field of the rotating frame and the electromagnetic field in the relativistic region \cite{a2,a1,com2}. 
Nevertheless, as we have seen, the rotating frame can be treated consistently within the usual framework of quantum mechanics, 
and it is shown that phase shifts to the first order are obtained {\it without any new hypothesis}. 

 Moreover, we directly obtained the Hamiltonian (\ref{H5}) for the uniform rotating frame from the Dirac equation with the spin connection in the low energy limit, 
 which is analogous to the Hamiltonian of the magnetic dipole in the electric field \cite{a6}, which gives rise to the Aharonov-Casher effect \cite{a-c}. 
 An analogous Hamiltonian ((3.18) in ref. \cite{a2}) in a gravitational field was obtained by one of us by considering the parallel transport of the wave function \cite{com3}.
We have in fact extended this Hamiltonian to the rotating frame. We shall discuss the further in a future paper. \\[12pt]

\noindent
{\Large {\bf Acknowledgments}}\\[12pt]
This work was supported in part by an NSF grant and an ONR grant. \\[12pt]

\noindent
{\Large {\bf Appendix: Conventions and Notations}}\\[12pt]
The metric is written as $g_{\mu \nu}=\eta _{\mu \nu} +h_{\mu \nu}$ where $\eta _{\mu \nu}={\rm diag} (+---)$.
Both indices $\mu,\nu$ and $a,b$ run over $0,1,2,3$, on the other hand $i,j,k$ run over $1,2,3$.

The vierbein $e^{\mu}_{\ a}$ and its inverse $e^{\ a}_{\mu}$ at each point which satisfy 
$g_{\mu \nu}=\eta _{ab}e^{\ a}_{\mu}e^{\ b}_{\nu}$ and $e^{\ a}_{\mu}e^{\mu}_{\ b}=\delta^a_{\ b}$. And they are used to connect latin 
indices and greek indices, for instance, $\gamma ^a =e^{\ a}_{\mu} \gamma ^{\mu}$ where $\gamma ^a$ and $\gamma ^{\mu}$ satisfy 
$\{ \gamma ^a ,\gamma ^b \}=2 \eta ^{ab}$ and $\{ \gamma ^{\mu} ,\gamma ^{\nu} \}=2 g^{\mu \nu}$ respectively. 
The Minkowski metric $\eta _{ab}$ and its inverse are used to lower and raise latin indices. 
In the weak field limit, the vierbein and its inverse can be written as $e^{\mu}_{\ a}=\delta^{\mu}_{\ a} -\frac 12 h^{\mu}_{\ a}$ and 
$e^{a}_{\ \mu}=\delta^{a}_{\ \mu} +\frac 12 h^{a}_{\ \mu}$, and we can check them to satisfy above properties to first order. 
$\Gamma ^{ab} _{\  \mu}$ in (\ref{cov}) are the Ricci rotation coefficients, and in the weak field limit
$\Gamma _{ab  \mu} =\frac 12 ( \partial _a h_{\mu b} -\partial _b h_{\mu a}) = - \Gamma _{ba  \mu} $. \\[12pt]

\noindent
{\Large {\bf Appendix 2: Reply to comments by G. Papini }}\\[12pt]
The issue of gauge invariance commented on by G. Papini \cite{papini} was discussed by one of us \cite{a7}. 
The Schr\"odinger equation in a weak gravitational field discussed by Papini \cite{dewitt}, 
 \begin{equation}  \tag{A.1}
i \frac{ \partial \psi}{\partial t} = [\frac{1}{2m} (p^i +m h_{0i})^2 + \frac 12 m h_{00}] \psi \label{schapp}
\end{equation}
would not violate gauge invariance if it is interpreted to be valid only in coordinate systems at rest with respect to the apparatus, i.e. the 
4-velocity field of the apparatus $t^{\mu}$ is proportional to $\delta ^{\ \mu}_{0}$ in all such coordinate systems. 
Under the transformation between any two such coordinate systems 
\begin{equation} \tag{A.2}
t^{' \mu } = \frac{\partial x^{' \mu}}{\partial x^{\nu}}t^{\nu}   \ , \label{trans} 
\end{equation}
with $t^{\mu} \propto  \delta ^{\ \mu}_{0}$ and $t^{' \mu} \propto  \delta ^{\ \mu}_{0}$ . 
For an infinitesimal coordinate transformation $x^{' \mu} = x^{\mu} +\xi ^{\mu} $, Eq.(\ref{trans}) implies $\xi^{i} _{\ ,0}=0,~ i=1,2,3$. 
Therefore the usual transformation $h_{\mu \nu} \rightarrow h_{\mu \nu} - \xi _{\mu ,\nu} -\xi_{\nu ,\mu}$ with this restriction 
transforms $G_{\mu} =(\frac 12 h_{00} , -h_{0i}) $ that is minimally coupled in (\ref{schapp}) according to $G_{\mu} \rightarrow G_{\mu} -\partial _{\mu} \xi_0$. 
Thus the transformation of $G_{\mu}$ is entirely analogous to the gauge transformation of the electromagnetic 4-vector potential $A_{\mu}$. 
This ensures that the Schr\"odinger equation (\ref{schapp}) and the phase shift $\delta \phi  =- \frac {m}{\hbar} \oint G_{\mu} dx^{\mu}$, 
which includes the Sagnac phase shift, are gauge invariant. 

Therefore, it is not necessary to require that $h_{00}$ and $h_{0i}$ are time independent as mentioned by Papini. 
In fact, the latter requirement does not help to preserve gauge invariance because it is possible for
 $\xi ^{i}_{\ ,0}$ to be time independent, yet non zero, in which case $h_{00}$ and $h_{0i}$ could be time independent in both coordinate systems; 
 yet Eq.(\ref{schapp}) and the phase shift would not be gauge invariant.
 
 Regarding Papini's question, we emphasize that, in our approximation, we neglect terms $v^2/c^2$, where $v$ is the velocity {\it relative to the rotating frame}. But we keep terms $\Omega^2/c^2$, because in a rotating coordinate system, $\Omega$ is a parameter determining the inertial fields
 (Coriolis and centrifugal fields). In other words, our low energy approximation is in the rotating coordinate system, and not in an inertial coordinate system.


\begin{thebibliography}{99}
\bibitem{com2} It should be remarked that if we use the Pauli equation instead of the Dirac equation (\ref{dirac2}) and plug (\ref{bmu}) into it, then we obtain 
$i \frac{\partial }{\partial t} \psi = \{ \frac {1}{2m}(\hat{\boldsymbol p} - m \boldsymbol {\cal A} )^2 
+m {\cal A}_0  -2 \boldsymbol \Omega  \cdot \hat{\boldsymbol S} \} \psi$ which has the twice factor in front of the spin term.
\bibitem{com3} See section 3 in ref. \cite{a2}. 
Note that the gauge used in this reference is 
different from ours, which explains the differece between the Hamiltonian ((3.18) in ref. \cite{a2}) and our Hamiltonian (\ref{H5}).
The former gauge in the absence of gravity, as in the present case, would pick out an inertial Minkowskian coordinates system, which is different from the 
rotating coordinates system used present pepar. 
 But it is straightforward to obtain the Hamiltonian for our case by using the method described there.
\end{thebibliography}
\end{document}